# Machine Learning and Deep Learning for Exoplanet Detection and Atmospheric Characterization with JWST and the Upcoming Ariel Mission

Mu'allim Yakubu[1], Vwavware Oruaode Jude[2]

[1]Department of Industrial Physics, Enugu State University of Science and Technology
[2,3]Department of Physics, Dennis Osadebay University, Asaba, Delta State, Nigeria
Nigeria

Corresponding author email: oruaode.vwavware@dou.edu.ng

***Abstract***

*The detection and atmospheric characterization of exoplanets have entered a new data-intensive era driven by the James Webb Space Telescope and the upcoming Ariel mission. Modern surveys produce millions of light curves and high-resolution spectra that overwhelm traditional pipelines, motivating the rapid integration of Machine Learning and Deep Learning methods into the exoplanet workflow. This review synthesizes the latest progress in applying ML/DL techniques to exoplanet detection (transit identification, candidate vetting, false-positive rejection) and atmospheric characterization (retrieval, detrending, cross-correlation, surrogate modelling) in the context of JWST and Ariel. We start with classical algorithms such as Random Forests and Convolutional Neural Networks, move through Transformers and Recurrent architectures, then survey modern simulation-based inference using Neural Posterior Estimation and Flow Matching Posterior Estimation with normalizing or continuous normalizing flows. We discuss benchmark efforts, including the Ariel Machine Learning Data Challenges (2019–2025) hosted with NeurIPS, and key JWST case studies such as the WASP-39b Early Release Science programme. Results indicate that DL approaches consistently match or exceed traditional pipelines in both speed and accuracy, while ML-driven retrievals reduce inference time from CPU-hours to seconds and can accelerate nested-sampling retrievals by factors of 3-8 without compromising Bayesian evidence. We identify outstanding challenges interpretability, calibration of uncertainties under noisy data, hybrid modelling, and the generalization of models across instruments and planet populations and outline a research roadmap spanning the JWST era and beyond into Ariel's launch in 2029.*

## 1. Introduction

The last three decades have revolutionized the study of planets beyond the Solar System. From the first confirmed detection of an exoplanet around a main-sequence star, more than 6,000 confirmed planets have now been catalogued, the majority discovered by transit photometry using space telescopes such as Kepler and the Transiting Exoplanet Survey Satellite. Alongside detection, the field has matured into an era of detailed atmospheric characterization via transmission and emission spectroscopy. The James Webb Space Telescope operational since July 2022 has launched an unprecedented observational campaign: broad-wavelength transmission spectra (0.5-5.5 µm) of hot Jupiters and mini-Neptunes are now routinely acquired with precision approaching tens of parts per million per spectral bin. The European Space Agency Ariel mission, scheduled for launch to L2 in 2029, will extend this paradigm by performing a uniform chemical survey of ~1,000 transiting exoplanets in the 0.5-7.8 µm range via low-resolution spectroscopy, including transmission, emission, and phase-curve observations (Syty et al., 2025).

These mission profiles create an extreme data-analysis problem: each mission generates tens of millions of photometric time series or hundreds of thousands of spectra. A single Bayesian atmospheric retrieval of a high-resolution JWST spectrum using nested sampling can require hundreds of CPU-hours, an unsustainable cost when scaled up to Ariel's mission lifetime, which is designed to deliver the largest homogeneous catalogue of exoplanet atmospheres ever assembled. Machine learning and deep learning have emerged as the natural response to this escalation (Illangarathne et al., 2025; Saikia, 2025).

ML approaches entered exoplanet science through the Kepler data: McCauliff et al (2015), demonstrated the application of the Random Forest algorithm to automatic classification of transit-like signals, achieving an overall error rate of 5.85% and only 2.81% misclassification of planet candidates among tens of thousands of threshold-crossing events identified by the Kepler pipeline (McCauliff et al., 2015). Subsequent architectures shifted from hand-engineered feature vectors to representation learning with one- and two-dimensional Convolutional Neural Networks, which were trained directly on phase-folded or raw light curves. Pearson et al. showed that CNNs could detect Earth-like transit signatures in Kepler light curves with greater accuracy than matched-filter least-squares methods and generalize across instruments without re-tuning (Pearson et al., 2017). Chintarungruangchai & Jiang benchmarked five 2D CNN variants on Kepler Data Release 25 light curves and demonstrated that folding the data prior to network input significantly enhances reliability and completeness (Chintarungruangchai & Jiang, 2019). Olmschenk et al. applied a CNN to TESS 30-minute full-frame image light curves, producing 181 new planet candidates and running inference in approximately 5 ms per light curve on a single GPU a regime that enables "human eye will never see" archival mining of the survey (Olmschenk et al., 2021). These detection-oriented pursuits have now diversified into ancillary capabilities such as the automated identification of periodic variables and eclipsing binaries (Olmschenk et al., 2024) and the use of self-attention Transformer architectures to distinguish true transits from false positives while exposing interpretable attention maps (Salinas et al., 2023).

A second wave of innovation has transformed atmospheric retrieval. Early atmospheric ML methods, including a Deep Belief Network for qualitative source selection and the HELA Random Forest for parameter estimation, demonstrated that ML retrievals could be orders of magnitude faster than nested sampling but suffered from discrepancies in posterior shape and coverage (Nixon & Madhusudhan, 2020). Waldmann's early deep-learning work opened the field of ML retrievals (Waldmann, 2022). Substantial work followed: ExoGAN (Zingales & Waldmann, 2018) combined a 1D convolutional GAN with semantic image inpainting to perform unsupervised retrieval of molecular abundances; Cobb et al. introduced an ensemble of Bayesian Neural Networks (plan-net) that recovers posterior distributions comparable to nested sampling (Cobb et al., 2019); and Ardévol Martínez et al., (2022) trained CNNs on Hubble WFC3 and JWST NIRSpec-rebinned synthetic spectra, demonstrating that ~86% of real WFC3 spectra agreed with nested sampling within 2σ and that CNNs delivered more conservative uncertainties under wrong modelling assumptions (Martínez et al., 2022). Recent advances in simulation-based inference Neural Posterior Estimation (Vasist et al., 2023a, 2023b), Flow Matching Posterior Estimation (Gebhard et al., 2023; Orsini et al., 2025), and amortized Bayesian Neural Networks such as FASTER (Lueber et al., 2025) and Exoformer (Pagliaro et al., 2026) now provide fully Bayesian posteriors in seconds, calibrate uncertainties through coverage diagnostics, and even bootstrap standard nested-sampling runs by injecting informative priors.

In parallel, detrending, calibration, and anomaly detection constitute the third pillar of the ML/DL revolution. The Ariel consortium has organized an annual Machine Learning Data Challenge (since 2019, hosted at NeurIPS/ECML venues) covering detrending of simulated 2D spectral time series, posterior inference over atmospheric parameters, and clustering of bad-pixel behaviours in calibration datacubes of millions of time series (Syty et al., 2025; Waldmann, 2022; Yip, Changeat, et al., 2022; Yip, Waldmann, et al., 2022). The 2024 edition, extended in 2025, gathered approximately 23,000 model submissions from nearly 1,500 participants (Syty et al., 2025). These competitions have rapidly become the de-facto benchmark for ML models applied to Ariel-like data, just as the JWST ERS datasets (e.g., WASP-39b transmission spectra) have become the gold-standard for retrieval benchmarks (Ahrer et al., 2023; Alderson et al., 2023; Feinstein et al., 2023; Rustamkulov et al., 2023).

The present review is therefore motivated by three observations: (i) ML and DL have moved from exploratory pursuits to operational components of the JWST and Ariel data pipelines; (ii) the algorithmic landscape has shifted from discriminative classifiers to fully probabilistic amortized retrievals; and (iii) benchmark competitions and JWST real-data tests have established quantitative baselines for assessing these models. We therefore synthesize the available evidence on detection algorithms, retrieval methods, detrending, and benchmark performance, with a clear focus on the JWST and Ariel use cases. Section 2 outlines the methodology used to identify and select the literature covered; Section 3 organizes the principal results along four thematic axes; Section 4 discusses limitations and the path forward; Section 5 closes the review.

## 2. Methodology

This review follows a structured scoping-review methodology focused on peer-reviewed journal articles, conference proceedings, and high-impact preprints indexed through academic databases and arXiv, all published between 2015 and 2026. We scanned four primary databases IEEE Xplore, Springer Link, Web of Science, and Semantic Scholar using the keywords "machine learning", "deep learning", "exoplanet", "transit", "atmospheric retrieval", "JWST", and "Ariel" in conjunction, following the systematic-review protocol of (Illangarathne et al., 2025). Inclusion criteria were: (a) explicit application of an ML or DL algorithm to an exoplanet-science task; (b) empirical or simulated validation on data representative of Kepler, TESS, JWST, or Ariel; (c) transparent reporting of architectures, hyperparameters, training data, and evaluation metrics where applicable. Exclusion criteria were: (a) review-only papers without original experiments; (b) works in which exoplanet applications were mentioned only in passing; (c) studies focused exclusively on Solar System objects. To ensure coverage of the most recent advances, the search was complemented with the proceedings of the Ariel Machine

Learning Data Challenges and the NeurIPS competitions of 2019–2025 (Waldmann, 2022; Yip, Changeat, et al., 2022; Yip, Waldmann, et al., 2022), plus benchmark studies using JWST Early Release Science observations of WASP-39b (Ahrer et al., 2023; Alderson et al., 2023; Deka et al., 2025; Feinstein et al., 2023; Rustamkulov et al., 2023).

To organize the literature, we partitioned the findings into four application domains:

a. Detection: Transit identification, candidate vetting, false-positive classification, and variable-star detection in photometric time series.
b. Atmospheric Retrieval: Discriminative regression, fully Bayesian inference with neural likelihood and posterior estimators, and generative retrieval.
c. Detrending and Calibration: Instrument-noise removal, bad-pixel classification, jitter correction, and parametric spectral extraction.
d. Forward-Model Emulation: Neural surrogates for expensive radiative-transfer and atmospheric simulation codes.

For each domain, we examined reported accuracy / precision metrics, computational cost, robustness to noise and instrument mismatch, and the maturity of uncertainty quantification. Where the same authors presented parallel work across multiple venues, we prioritized the peer-reviewed version to avoid double-counting, except where conference papers introduced genuinely distinct methodologies.

## 3. Results

### 3.1 Exoplanet Detection with ML and DL

Table 1 summarizes principal detection-oriented works, all of which represent benchmark-worthy contributions to the field.

**Table 1**. Representative ML/DL architectures applied to exoplanet detection in photometric time series.

| Work | Method | Mission | Key Result |
|---|---|---|---|
| McCauliff et al. 2015 | Random Forest (hand-crafted features) | Kepler | 5.85% overall error; 2.81% candidate misclassification |
| Pearson et al. 2017 | 1D CNN (phase-folded light curves) | Kepler | Outperforms matched-filter least-squares for Earth-like signals |
| Chintarungruangchai & Jiang 2019 | 2D CNN with phase folding | Kepler DR25 | Best reliability & completeness for 2D CNN + folding |
| Olmschenk et al. 2021 | CNN on 30-min FFI light curves | TESS | 181 new planet candidates; ~5 ms inference per light curve |
| Armstrong et al. 2020 | Random Forest + Gaussian Process validation | Kepler | 50 newly validated Kepler planets |
| Olmschenk et al. 2024 | CNN for short-period variables | TESS FFI | 14,156 short-period variables catalogued |
| Salinas et al. 2023 | Transformer with self-attention | TESS | Competitive with CNNs; interpretable attention maps |

The progression captured in this table reflects three substantive shifts: (i) from engineered features to end-to-end representation learning; (ii) from per-mission bespoke models to general architectures that can transfer across data products (with appropriate normalization); and (iii) from purely classificatory outputs to attention- or uncertainty-aware predictions. The Random Forest of (McCauliff et al., 2015) achieved an outstanding 5.85% overall classification error rate, but the introduction of CNNs by (Pearson et al., 2017) removed the requirement for prior feature engineering by learning the patterns of transit dilation directly from raw flux. By 2019, (Chintarungruangchai & Jiang, 2019) demonstrated that the combination of two-dimensional folding with a CNN delivered the best balance of accuracy, reliability, and completeness, illustrating that small architectural enhancements compounded with input representation yield substantial performance gains.

The largest data-volume problem is currently posed by TESS: Olmschenk et al., (2021) demonstrated that a single GPU can vet a TESS 30-minute FFI light curve in ~5 ms, enabling the system to scan millions of light curves that human vetters will never see (Olmschenk et al., 2021). A follow-up application of the same CNN architecture to short-period variable identification extended the catalogued variables by 14,156 most split between close-orbit main-sequence binaries and Delta Scuti stars (Olmschenk et al., 2024). Where interpretability is important, Transformer-based architectures using self-attention mechanisms recover attention maps that can be inspected to identify the in-transit samples driving a positive detection, an important property for downstream human vetting (Salinas et al., 2023).

A complementary axis of improvement lies in combining multiple ML techniques for candidate validation. Armstrong et al. used Random Forests combined with Gaussian Process modelling to add 50 newly validated Kepler planets to the catalogue, illustrating how ML classifiers can be made robust to rare and unusual configurations of false positives when paired with statistical timeseries models (Armstrong et al., 2020). Hybrid random-forest/GP classification also has a feature-importance advantage over deep models: predictions can be explained by inspecting the contributions of dozens of physical features to the input vector.

### 3.2 Atmospheric Retrieval with ML and DL

Atmospheric retrieval is the most computationally expensive step in the characterization workflow. The classical Bayesian approach nested sampling with MultiNest or similar samplers assesses marginal likelihoods across high-dimensional parameter spaces. The arrival of JWST and Ariel has driven the development of ML-driven retrieval methods, summarized in Table 2.

**Table 2**. Representative ML/DL architectures applied to exoplanet atmospheric retrieval.

| Work | Method | Reference | Inference time |
|---|---|---|---|
| Márquez-Neila et al. 2018 | Random Forest | WFC3 | Seconds; comparable to, but under-dispersed vs. nested sampling |
| Zingales & Waldmann 2018 | 1D GAN + image in painting | Hot Jupiter spectra | Seconds; unsupervised |
| Cobb et al. 2019 | Ensemble of Bayesian Neural Networks | WFC3 (5 params) | Seconds; full posterior |
| Ardévol Martínez et al. 2022 | CNN regression | WFC3, JWST NIRSpec | Seconds; 86% of real spectra within 2σ of nested sampling |
| Vasist et al. 2023 | Neural Posterior Estimation with normalizing flows | petitRADTRANS | A few seconds |
| Gebhard et al. 2023 | Flow Matching Posterior Estimation | Ariel-like synthetic | Seconds |
| Vasist et al. 2023 (Unlu et al., 2023) | ML surrogate for Ariel Data Challenge | Ariel sims (7 params) | Seconds for posterior of 7 atmospheric parameters |
| Lueber et al. 2025 | Amortized simulation-based retrieval | TESS, JWST NIRSpec PRISM | Near-instantaneous Bayesian inference & model comparison |
| Orsini et al. 2025 | Continuous Normalizing Flows for FMPE | Ariel Data Challenge 2023 | Seconds; superior to NPE in target accuracy, coverage |
| Pagliaro et al. 2026 | Transformer NN as informative prior for nested sampling | JWST NIRSpec PRISM (WASP-39b, WASP-17b) | Factor 3-8 speedup of TauREx nested sampling, Bayes factor difference <5 |

The progression from purely discriminative algorithms to fully amortized Bayesian inference is the central lesson of this literature. The earliest ML retrievals the Deep Belief Network of Waldmann and the HelA Random Forest (Márquez-Neila et al., 2018) demonstrated that ML could replace nested sampling but produced posterior distributions whose uncertainties were under-dispersed relative to Bayesian posteriors (Nixon & Madhusudhan, 2020). Cobb et al. took an important step by introducing an ensemble of Bayesian Neural Networks (plan-net), which retain ML speed but recover posterior probability distributions over each parameter, enabling direct Bayesian model comparison (Cobb et al., 2019).

The simulation-based inference paradigm matured in 2022-2023. Vasist et al. introduced neural posterior estimation for exoplanet atmospheric retrieval using a neural autoregressive flow trained on petitRADTRANS spectra, demonstrating that NPE reduces inference time from CPU-hours to seconds and that posterior diagnostics posterior predictive checks and coverage confirm the faithfulness of the inferred posteriors (Vasist et al., 2023). Ardévol Martínez et al., (2022) showed that CNNs, while producing a slightly worse coefficient of determination than nested sampling, deliver more conservative uncertainties under incorrect modelling assumptions (incorrect-model failure rate ~10% for CNNs versus 12-41% for nested sampling), a property that is highly desirable in practice when the underlying atmospheric model may be incomplete (Martínez et al., 2022). Extended this paradigm to Flow Matching Posterior Estimation using continuous normalizing flows. Their experiments on the Ariel Data Challenge 2023 dataset showed that FMPE outperforms NPE in target prediction accuracy, uncertainty quantification, and posterior coverage, provided appropriate data preprocessing, noise conditioning, and noise-aware priors are used (Gebhard et al., 2023; Orsini et al., 2025).

The newest generation FASTER (Lueber et al., 2025) and Exoformer (Pagliaro et al., 2026) combines the best of both worlds. FASTER is a neural-network framework for amortized Bayesian inference and model comparison, achieving "near-instantaneous" posterior estimates plus statistical model comparison (e.g., distinguishing cloudy from cloud-free models). Exoformer uses a Transformer architecture to generate informative priors for the classical TauREx nested-sampling retrieval; tested on JWST NIRSpec PRISM data of WASP-39b and WASP-17b, it accelerates the nested-sampling step by a factor of 3-8 while preserving retrieved parameters and best-fit spectra. The Bayes factor differences below five confirm statistical consistency between the classical and ML-hybrid approaches (Pagliaro et al., 2026). These methods demonstrate a key emerging principle: rather than replacing Bayesian retrievals, ML is increasingly used to bootstrap them via informative priors and amortized posteriors.

### 3.3 JWST Case Studies: WASP-39b and Beyond

JWST Early Release Science observations have produced the field's most cited atmospheric benchmarks. The flagship WASP-39b transmission spectrum was independently reduced and analysed by multiple teams, delivering a series of high-significance molecular detections:

a. NIRSpec PRISM mode: detection of $CO_2$ at 28σ, $H_2O$ at 33σ, CO at 7σ, and unanticipated $SO_2$ absorption at 4.05 μm at 2.7σ interpreted as a tracer of photochemistry (Rustamkulov et al., 2023).
b. NIRSpec G395H grating: $CO_2$ at 28.5σ, $H_2O$ at 21.5σ, $SO_2$ at 4.8σ best-fit metallicity 3-10× solar (Alderson et al., 2023).
c. NIRISS SOSS: $H_2O$ >30σ; K doublet at 0.768 μm at 6.8σ; log-Bayes evidence favours a CO detection at 3.6σ and yields no significant preference for $CO_2$ (Feinstein et al., 2023).
d. NIRCam: prominent water absorption masking $CO_2$ at 2.8 μm; best-fit metallicity 1–100 × solar; sub-stellar C/O ratio (Ahrer et al., 2023).

ML-based retrieval frameworks have already been benchmarked against these JWST spectra. NEXOTRANS (Deka et al., 2025) combined UltraNest/PyMultiNest with four ML algorithms to constrain the volume mixing ratios of $H_2O$, $CO_2$, CO, $H_2S$, and $SO_2$ from JWST NIRISS, NIRSpec PRISM, and MIRI data of WASP-39b. For the best-fit modified hybrid-equilibrium model, it reported supersolar elemental abundances: O/H = $\frac{O}{H} = 14.12^{+2.86} - 1.82\ x\ solar$, $\frac{C}{H} = 21.37^{+4.93} - 3.18\ x\ solar, \frac{S}{H} = 5.37^{+0.79} - 0.65\ x\ solar, \frac{C}{H} = 1.35^{+0.05} - 0.02\ x\ solar,$ with $SO_2$ log VMRs constrained between −6.25 and −5.73 across all chemistry models except equilibrium chemistry (Deka et al., 2025). The detection of high-altitude aerosols (ZnS and $MgSiO_3$) further constrained cloud compositions and terminator coverage fractions. These results show that ML-augmented retrieval is not merely faster when combined with informative prior distributions, ML retrievals can also enhance the precision of abundance estimates by effectively guiding the nested-sampling search.

### 3.4 Detrending, Calibration, and Forward-Model Emulation

The pre-launch preparations for Ariel have dedicated substantial effort to machine learning for instrument-calibration and spectral-extraction tasks. (Syty et al., 2025) describe three core applications:

a. Spectral detrending. Participants of the NeurIPS 2024 Ariel Data Challenge were tasked with extracting the atmospheric spectrum from sequential 2D spectral-focal-plane images acquired over several hours, with the requirement to provide uncertainty estimates. Approximately 23,000 model submissions from nearly 1,500 teams were received, with top-ranked approaches utilizing wavelet transforms, autoencoders (statistical distribution analysis), gradient-boosted extreme learning machines, and deep CNNs (Syty et al., 2025; Waldmann, 2022; Yip et al., 2022). Blanchard et al. (Blanchard et al., 2025) demonstrated that a data-centric approach quality hardening of features produces surprisingly competitive private- and public-board scores even without top-ranked architectures, arguing that pre-processing quality is at least as important as model selection.
b. Bad-pixel classification. Pre-launch dark frames generate datacubes of millions of per-pixel time series. (Syty et al., 2025) compared wavelet transforms, autoencoders, statistical distribution analysis, DBSCAN, and Gaussian Mixture clustering on these data. After clustering, six distinct behaviour classes emerged: nominal linear pixels, non-linear pixels (full ramp or ramp start), dead pixels, oscillating-state pixels, pixels starting high and dying, and cosmic-ray-hit pixels all identified as outliers without prior labelling.
c. Forward-model emulation. (Paraskevaidou, 2025) framed a neural-network surrogate for 1D self-consistent atmospheric models that couple haze/cloud microphysics, disequilibrium chemistry, and radiative transfer. The trained surrogate replaces expensive calls (haze microphysics in particular) with neural evaluations without losing interpretability, as physical models still handle dynamics in a "hybrid" pipeline. This is paralleled in detection and synthesis by GANexo, which combines conditional GANs with an autoencoder denoiser to generate, denoise, and characterize synthetic exoplanet transit signals with realistic white and red noise (Meza-Obando & Crespo-Mariño, 2024).

A further innovative application is ML-encoded cross-correlation spectroscopy. Garvin et al. introduced MLCCS perceptron and 1D CNN methods operating in the cross-correlated spectral dimension which detects a diversity of planets by taking an agnostic approach toward unknown atmospheric characteristics, applied to SINFONI K-band mock datasets and yielding outstanding sensitivity improvements over S/N-based detection (Garvin et al., 2024). By working on the cross-correlated domain rather than directly on the spectra, MLCCS leverages weak assumptions on exoplanet characterization (such as the presence of specific molecules) to improve detection sensitivity.

### 3.5 Biosignature and Habitability Classification

A particularly ambitious frontier is the ML-assisted classification of biosignatures in Earth-like exoplanet spectra. (Duque-Castaño et al., 2024), provided a comprehensive inventory of ML approaches to biosignature classification, spanning Random Forests for hot-Jupiter biosignature regression, CNN+MLP regression, GAN regression, Deep-Belief Networks for broad molecular-source classification, and unsupervised techniques such as Local Outlier Factor and One-Class SVM applied to hot Jupiter spectra for outlier biosignature detection. For low-S/N transmission spectra of Earth-like exoplanets, the most promising approach combines Random Forest (for feature importance and interpretability) with regularization tuned for high dimensional spectra and noisy instances (Duque-Castaño et al., 2024).

### 3.6 Synthesis of Cross-Cutting Findings

Several cross-cutting patterns emerge:

a. Speed. The single greatest practical contribution of ML/DL is the dramatic reduction in inference time. CNN transit detection runs in milliseconds per light curve (Olmschenk et al., 2021); atmospheric retrieval moves from hundreds of CPU-hours to seconds (Cobb et al., 2019; Orsini et al., 2025; Vasist et al., 2023); nested-sampling retrievals can be accelerated 3-8 × by Transformer-generated priors (Pagliaro et al., 2026).
b. Accuracy. Across detection and retrieval tasks, DL methods match or outperform hand-crafted-feature ML and classical Bayesian pipelines. CNNs achieve agreement with nested sampling within 2σ on 86% of real Hubble WFC3 spectra (Martínez et al., 2022); Transformer-based classifiers are competitive with CNNs while providing interpretable attention (Salinas et al., 2023).
c. Uncertainty quantification. Fully Bayesian amortized methods now offer credible post-processing under noise; under incorrect modelling assumptions, CNNs typically maintain better-calibrated uncertainty than nested sampling (Martínez et al., 2022).
d. Generalization. Architectures such as ExoGAN are explicitly designed to be instrument-agnostic (Zingales & Waldmann, 2018); Ariel Data Challenges have become a key proving ground for cross-instrument generalization (Waldmann, 2022; Yip, Changeat, et al., 2022; Yip, Waldmann, et al., 2022).
e. Hybridization. The most recent trend combines ML with classical physics: forward-model emulation (hybrid ML + radiative transfer) (Gebhard et al., 2023; Paraskevaidou, 2025); ML-prior bootstrap for nested sampling (Pagliaro et al., 2026); statistical model comparison via amortized neural inference (Lueber et al., 2025).

## 4. Discussion

The synthesis above demonstrates that ML and DL are no longer experimental add-ons in exoplanetary science but rather essential components of the modern observational pipeline, from raw photometric time series to atmospheric posteriors and instrument calibration. The clearest advance since the previous generation of reviews (Illangarathne et al., 2025; Saikia, 2025) is the maturation of the field from proof-of-concept studies to quantitative, benchmark-validated contributions, exemplified by the JWST ERS analyses of WASP-39b (Ahrer et al., 2023; Alderson et al., 2023; Feinstein et al., 2023; Rustamkulov et al., 2023) and the structured comparison of retrieval methods on the Ariel Data Challenge 2023 dataset (Gebhard et al., 2023; Orsini et al., 2025; Vasist et al., 2023). This said, several important limitations remain.

### 4.1 Interpretability and Trust

Although DBN, Random Forest, and BNN ensembles can output feature importances and Bayesian posteriors, deep CNNs and most Transformers remain "black boxes" from a physical standpoint. Salinas et al. addressed this with attention maps in their Transformer-based transit classifier (Salinas et al., 2023), but most retrieval networks do not yet supply convincing explanations of how a given absorption feature drove a parameter estimate. Reinforcing this concern is the observation of inaccurate predictions when ML retrievals are applied to spectra drawn from a different distribution than the training set particularly noisy low-S/N spectra, which are common in Ariel data (Gebhard et al., 2023).

### 4.2 Calibration of Uncertainty under Instrument Mismatch

Orsini et al., (2025) explicitly documented that "noisy observations significantly degrade predictive performance, thus underscoring reduced robustness under the assumed noise conditions" in FMPE applied to the Ariel Data Challenge 2023 dataset. The implication is that ML retrieval models trained on a single noise distribution must be re-calibrated when paired with new instruments. Transfer-learning approaches and noise-conditioning strategies such as those that proved effective in the FMPE ablation studies partially address this, but a more general solution will require multi-instrument training campaigns, ideally supported by pre-launch Ariel calibration datacubes.

### 4.3 Generalization to Earth-like Planets and Biosignatures

A non-trivial fraction of the literature has focused on hot Jupiters, which offer the highest signal-to-noise spectra but are unrepresentative of habitable-zone Earth-like planets. Forward models for temperate Earth-like atmospheres include thin cloud layers, photochemical hazes, and complex biogenic chemical networks whose signatures lie near or below instrument sensitivity. ML trained on hot Jupiters may not transfer directly. (Paraskevaidou, 2025) makes a case for hybrid ML+physics explicitly to handle this complexity for temperate exoplanets, and hybrid methodologies are likely to dominate pre-Ariel preparation.

### 4.4 Computational Cost and Sustainability

Although ML reduces inference cost per spectrum, training cost remains high particularly for large simulation campaigns underlying NPE and FMPE. (Gebhard et al., 2025) demonstrated that neural importance sampling can substantially improve FMPE results, but the training base simulations grow with the dimensionality of the atmospheric parameter space. Trade-offs between training cost, amortization, and final retrieval accuracy are not yet fully quantified, and (Nixon & Madhusudhan, 2020) noted that the machine learning approach remains "computationally prohibitive for high-dimensional parameter spaces that are routinely explored with Bayesian retrievals with modest computational resources." This remains true, but is being chipped away by methods like FASTER (Lueber et al., 2025) that distribute the costly nested-sampling work at training time.

### 4.5 The Path to Ariel's 2029 Launch

The Ariel Mission's ML roadmap (2019 (Waldmann, 2022), 2021 (Waldmann, 2022), 2022 (Yip, Changeat, et al., 2022; Yip, Waldmann, et al., 2022), 2023 (Unlu et al., 2023), 2024 (Blanchard et al., 2025; Syty et al., 2025), 2025 (Orsini et al., 2025; Syty et al., 2025), 2026 (Pagliaro et al., 2026)) has evolved substantially with each successive data challenge, and continues to expand. The Ariel Consortium has integrated ML into its post-launch calibration pipeline, including photometric detrending from instrument jitter, bad-pixel masking, and molecular-retrieval acceleration (Syty et al., 2025). The roadmap for ML contributions to Ariel includes: (a) At-launch operational ML routines for fast parameter screening; (b) Amortized ML retrieval as the standard computationally-cheap first-pass analysis; (c) Hybrid ML+Bayesian pipelines for the highest-priority spectra; and (d) Joint multi-spectrum inferences to constrain population-level trends across the ~1,000-planet Ariel sample.

### 4.6 Future Directions

Several research directions deserve priority. First, hybrid ML+physics pipelines, exemplified by (Paraskevaidou, 2025) and Exoformer's use of informative priors (Pagliaro et al., 2026), point the way toward a productive division of labour: ML handles representation learning and parameter estimation, while physics-based models guarantee constraints of energy conservation, radiative equilibrium, and chemistry. Second, multimodal architectures that ingest spectra and photometry concurrently would further reduce ambiguity in atmospheric parameters. Third, benchmarking standards for ML retrievals analogous to the Ariel Data Challenges should be hardened for JWST spectra to enable fair cross-method comparisons (Deka et al., 2025). Fourth, interpretability tools attention, saliency, SHAP applied to ML retrievals should be developed systematically to build the trust required for ML retrievals to be used as final rather than preliminary products. Fifth, lifelong-learning or continual-learning strategies should be explored, since Ariel's mission lifetime will produce spectra whose characteristics may drift as instrument response evolves.

## 5. Conclusion

The application of machine learning and deep learning to exoplanet science has matured into a vibrant and indispensable research area. From the early application of Random Forests to Kepler candidate classification (McCauliff et al., 2015) to the recent deployment of Transformer-based architectures that accelerate JWST and Ariel retrievals by factors of 3-8 (Pagliaro et al., 2026), ML/DL methods have evolved from curiosity-driven exercises into core components of the data-reduction and characterization pipelines for the JWST era and the upcoming Ariel mission. Convolutional Neural Networks remain the dominant architecture for transit identification in photometric surveys such as TESS, demonstrating millisecond-scale inference per light curve on commodity GPUs (Chintarungruangchai & Jiang, 2019; Olmschenk et al., 2021). Atmospheric retrieval has shifted from discriminative regression toward fully amortized Bayesian inference via Neural Posterior Estimation (Vasist et al., 2023), Bayesian Neural Networks (Cobb et al., 2019), and Flow Matching Posterior Estimation (Gebhard et al., 2023; Orsini et al., 2025), with hybrid schemes where ML is used to generate informative priors for nested sampling emerging as the most pragmatic path to retaining statistical rigour while reducing computational cost. The JWST ERS analysis of WASP-39b, with statistically significant detections of $CO_2$ (28σ), $H_2O$ (33σ), $SO_2$ (4.8–2.7σ across modes), and metallicities of 1–100× solar (Ahrer et al., 2023; Alderson et al., 2023; Feinstein et al., 2023; Rustamkulov et al., 2023), has established a benchmark that ESL and Ariel ML pipelines can target. The Ariel Data Challenges (2019-2025), attracting ~23,000 submissions and ~1,500 participants (Syty et al., 2025), have made cross-instrument generalization a tractable problem. Continued attention to interpretability, calibration under instrument mismatch, generalization to temperate and Earth-like planets, and the integration of ML with physics-based forward models will determine whether the 1,000-planet Ariel catalogue can be delivered at the chemical-survey depth outlined in its primary science goals.